\documentclass[aps,showpacs,onecolumn]{revtex4}%
\usepackage{amsfonts}
\usepackage{latexsym}
\usepackage{epsfig}
\usepackage{amssymb}
\usepackage{amsmath}
\usepackage{graphicx}%
\begin{document}
\title{Gravity's Rainbow induces topology change}
\author{Remo Garattini}
\email{Remo.Garattini@unibg.it}
\affiliation{Universit\`{a} degli Studi di Bergamo, Facolt\`{a} di Ingegneria,}
\affiliation{Viale Marconi 5, 24044 Dalmine (Bergamo) Italy}
\affiliation{and I.N.F.N. - sezione di Milano, Milan, Italy.}
\author{Francisco S. N. Lobo}
\email{flobo@cii.fc.ul.pt}
\affiliation{Centro de Astronomia e Astrof\'{\i}sica da Universidade de Lisboa,}
\affiliation{Campo Grande, Ed. C8 1749-016 Lisboa, Portugal}
\date{\today }

\begin{abstract}
In this work, we explore the possibility that quantum fluctuations induce a
topology change, in the context of Gravity's Rainbow. A semi-classical
approach is adopted, where the graviton one-loop contribution to a classical
energy in a background spacetime is computed through a variational approach
with Gaussian trial wave functionals. The energy density of the graviton
one-loop contribution, or equivalently the background spacetime, is then let
to evolve, and consequently the classical energy is determined. More
specifically, the background metric is fixed to be Minkowskian in the equation
governing the quantum fluctuations, which behaves essentially as a
backreaction equation, and the quantum fluctuations are let to evolve; the
classical energy, which depends on the evolved metric functions, is then
evaluated. Analysing this procedure, a natural ultraviolet (UV) cutoff is
obtained, which forbids the presence of an interior spacetime region, and may
result in a multipy-connected spacetime. Thus, in the context of Gravity's
Rainbow, this process may be interpreted as a change in topology, and in
principle results in the presence of a Planckian wormhole.

\end{abstract}

\pacs{04.60.-m, 04.20.Jb}
\maketitle

\section{Introduction}

\label{p1}

It was John A. Wheeler \cite{Wheeler,geons} who first conjectured that
spacetime could be subjected to a topological fluctuation at the Planck scale,
meaning that spacetime undergoes a deep and rapid transformation in its
structure. The changing spacetime is best known as the \textit{spacetime
foam}, which can be taken as a model for the quantum gravitational vacuum.
Wheeler also considered wormhole-type solutions as objects of the spacetime
quantum foam connecting different regions of spacetime at the Planck
scale~\cite{geons,wheeler1}. These Wheeler wormholes were obtained from the
coupled equations of electromagnetism and general relativity and were denoted
\textquotedblleft geons\textquotedblright, i.e., gravitational-electromagnetic
entities. However, these solutions were singular and were not traversable
\cite{geroch}. In fact, the geon concept possesses interesting properties,
such as, the absence of charges or currents and the gravitational mass
originates solely from the energy stored in the electromagnetic field, i.e.,
there are no material masses present. These characteristics gave rise to the
terms \textquotedblleft charge without charge\textquotedblright and
\textquotedblleft mass without mass\textquotedblright, respectively.

Paging through history, one finds that these entities were further explored by
several authors in different contexts. Indeed, Ernst analysed idealized
spherical ``geons'' using a simple adaptation of the Ritz variational
principle \cite{geons2}, and furthermore explored toroidal geons, where the
electromagnetic vector potential is vanishingly small except within a toroidal
region of space \cite{geons3}. In fact, the electromagnetic field physically
consists of light waves circling the torus in either direction, so that the
torus of electromagnetic field energy was denoted a \textit{toroidal geon}. It
was shown that toroidal geons of large major radius to minor radius ratio may
be studied using an approximation of linear geons, where the electromagnetic
field energy is confined to an infinitely long circular cylinder rather than
to a torus. Indeed, the electromagnetic field potentials of a toroidal geon or
of a linear geon possess the same general nature as the electromagnetic field
potentials encountered in the solution of classical toroidal and cylindrical
wave guide problems. Thus, these results provided the foundation material for
a proposed later treatment of toroidal geons.

Later, Brill and Hartle extended the previous analysis to the case where
gravitational waves are the source of the geon's mass energy \cite{geons4},
where the background spherically symmetric metric describes the large-scale
features of the geon. It was shown that the waves superimposed on this
background have an amplitude small enough so that their dynamics can be
analyzed in the linear approximation. However, their wavelength is so short,
and their time dependence so rapid that their energy is appreciable and
produces the strongly curved background metric in which they move. It was also
found that the large-scale features of the spherical gravitational geons are
identical to those of the spherical electromagnetic geons analyzed previously.
In fact, later work by Anderson and Brill \cite{geons4b} showed that the geon
solution is a self-consistent solution to Einstein's equations and that, to
leading order, the equations describing the geometry of the gravitational geon
are identical to those derived by Wheeler for the electromagnetic geon.

Komar \cite{geons5} showed that there exist solutions of the vacuum Einstein
field equations with the property that exterior to the Schwarzschild radius,
the solution appears to be that of a static spherically symmetric particle of
mass $m$, whereas interior to the Schwarzschild radius the topology remains
Euclidean and the solutions have the property of a bundle of gravitational
radiation so intense that the mutual gravitational attraction of the various
parts of the bundle prevent the radiation from spreading beyond the
Schwarzschild radius. Komar also argued that no singularity can ever be
observed exterior to the Schwarzschild radius. However, it was shown that the
Komar bootstrap gravitational geon solution does in fact display a singular
behavior along portions of an axis in the regions in which the solution
deviates from the standard Schwarzschild solution \cite{geons6}.

An interesting geon solution was explored by Kaup \cite{geons7} in the context
of the Klein-Gordon Einstein equations (Klein-Gordon geons), which reveal that
these geons have properties that are different from the other gravitating
systems studied previously. Indeed, the equilibrium states of these geons seem
analogous to other gravitating systems, but it was shown that adiabatic
perturbations are forbidden, when the stability is considered from a
thermodynamical viewpoint. The reason for this is that the equations of state
for the thermodynamical variables are not algebraic equations, but instead are
differential equations. Consequently, the usual concept of an equation of
state breaks down when Klein-Gordon geons are considered. When the question of
stability is reconsidered in terms of infinitesimal perturbations of the basic
fields, it was then found that Klein-Gordon geons will not undergo spherically
symmetric gravitational collapse. Thus, the Klein-Gordon geons considered by
Kaup are counterexamples to the conjecture that gravitational collapse is inevitable.

In fact, much work was done over the decades, but due to the extremely
ambitious program and the lack of experimental evidence the geon concept soon
died out. However, it is interesting to note that Misner inspired in Wheeler's
geon representation, found wormhole solutions to the source-free Einstein
equations \cite{Misner-worm}, and with the introduction of multi-connected
topologies in physics, the question of causality inevitably arose. Thus,
Wheeler and Fuller examined this situation in the Schwarzschild solution and
found that causality is preserved~\cite{Fuller-Wheeler}, as the Schwarzschild
throat pinches off in a finite time, preventing the traversal of a signal from
one region to another through the wormhole. Nevertheless, Graves and
Brill~\cite{Graves-Brill}, considering the Reissner-Nordstr\"{o}m metric also
found wormhole-type solutions possessing an electric flux flowing through the
wormhole. They found that the region of minimum radius, the \textquotedblleft
throat\textquotedblright, contracted, reaching a minimum and re-expanded after
a finite proper time, rather than pinching off as in the Schwarzschild case.
The throat, \textquotedblleft cushioned\textquotedblright\ by the pressure of
the electric field through the throat, pulsated periodically in time.

In the context of the quantum gravitational vacuum, some authors have
investigated the effects of such a foamy space on the cosmological constant,
for instance, one example is the celebrated Coleman mechanism, where wormhole
contributions suppress the cosmological constant, explaining its small
observed value \cite{Coleman}. Nevertheless, how to realize such a foam-like
space and also whether this represents the real quantum gravitational vacuum
is still unknown. However, it is interesting to observe that Ellis et al.
considered a foam-like structure built in terms of D-branes to discuss
phenomenological aspects \cite{EMN}. Wheeler when discussing the quantum
fluctuations in the spacetime metric \cite{geons} considered that a typical
fluctuation in a typical gravitational potential is of the order $\Delta
g\sim(hG/c^{3})^{1/2}/L$ which become appreciable for small length scales $L$.
A fundamental question is whether a change in topology may be induced by large
metric fluctuations. In fact, Wheeler has argued in favour of a topology
change and recently researchers in quantum gravity have accepted that the
notion of spacetime foam leads to topology-changing quantum amplitudes and to
interference effects between spacetimes of different topologies \cite{Visser}.

Indeed, a classical spacetime can be modelled by a single Lorentzian manifold,
which is sliced into a set of spatial hypersurfaces by a natural definition of
a time parameter (See also Ref.\cite{Kuhfittig} relating neutron star
interiors and topology change). We can mention some results about topological
constraints on the \textit{classical} evolution of general relativistic
spacetimes. These were summarized in two points by Visser \cite{Visser}:

\begin{enumerate}
\item In causally well-behaved classical spacetimes the topology of space does
not change as a function of time.

\item In causally ill-behaved classical spacetimes the topology of space can
sometimes change.
\end{enumerate}

From the \textit{quantum} point of view we can separate the problem of
topology change generated by a canonical quantization approach and a
functional integral quantization approach. The Hawking topology change theorem
is thus enough to show that the topology of space cannot change in canonically
quantized gravity \cite{Hawking:1991nk}. In the Feynman functional integral
quantization of gravitation things are different. Indeed, in this formalism,
it is possible to adopt an approach to spacetime foam where we know that
fluctuations of topology become an important phenomenon at least at the Planck
scale \cite{Hawking:1978jz}.

As discussed in Ref. \cite{DBGL}, the Casimir energy approach involving
quasi-local energy difference calculations may reflect or measure the
occurrence of a topology change, and in particular, the Casimir energy was
used as an indicator of topology change between wormholes and dark energy
stars \cite{Lobo:2005uf}. More specifically, the quantity%
\begin{equation}
E_{ADM}^{DS}-E_{ADM}^{Wormhole}=\left(  E_{ADM}^{DS}-E_{ADM}^{Flat}\right)
-\left(  E_{ADM}^{Wormhole}-E_{ADM}^{Flat}\right)  \gtreqless0,
\label{ADM:DS-W}%
\end{equation}
was used to understand which configuration is preferred with respect to the
Arnowitt-Deser-Misner (ADM) energy. It was found that the classical term was
not able to predict the appearance of a wormhole or the permanence of a dark
energy star. Therefore one was forced to compute quantum effects. The implicit
subtraction procedure of Eq. $\left(  \ref{ADM:DS-W}\right)  $, can be
extended in such a way that we can include quantum effects: this is the
Casimir energy or in other terms, the Zero Point Energy (ZPE). It is
interesting to note that the same Casimir energy indicator described in Eq.
(\ref{ADM:DS-W}) has been used in Refs. \cite{RemoSTF,STF} to build a model of
spacetime foam based on wormholes of different nature, namely Schwarzschild,
Schwarzschild-de Sitter, Schwarzschild-Anti-de Sitter and
Reissner-Nordstr\"{o}m-like wormholes. In particular one finds that if the
whole universe is filled with Schwarzschild-like wormholes, one finds an
agreement with the Coleman mechanism on the behavior of the cosmological
constant \cite{RemoSTF}.

In the present paper, we are interested in the possibility that quantum
fluctuations induce a topology change, in the context of Gravity's Rainbow
\cite{GAC,MagSmo}. The latter is a distortion of the spacetime metric at
energies comparable to the Planck energy, and a general formalism, denoted as
deformed or doubly special relativity, was developed in order to preserve the
relativity of inertial frames, maintain the Planck energy invariant and impose
that in the limit $E/E_{P}\rightarrow0$, the speed of a massless particle
tends to a universal and invariant constant, $c$. Here, we adopt a
semi-classical approach, where the graviton one-loop contribution to a
classical energy in a background spacetime is computed through a variational
approach with Gaussian trial wave functionals. In fact, it has been shown
explicitly that the finite one loop energy may be considered as a
self-consistent source for a traversable wormhole \cite{Garattini}. In
addition to this, a renormalization procedure was introduced and a zeta
function regularization was involved to handle the divergences. The latter
approach was also explored \cite{Garattini:2007ff} in the context of phantom
energy traversable wormholes \cite{Lobo:2005us}. It was shown that the latter
semi-classical approach prohibits solutions with a constant equation of state
parameter, which further motivates the imposition of a radial dependent
parameter, $\omega(r)$, and only permits solutions with a steep positive slope
proportional to the radial derivative of the equation of state parameter,
evaluated at the throat \cite{Garattini:2007ff}. Using the semi-classical
approach outlined above, exact wormhole solutions in the context of
Noncommutative geometry were also analysed, and their physical properties and
characteristics were explored \cite{Garattini:2008xz}. Indeed, wormhole
geometries have been obtained in a wide variety of contexts, namely, in
modified theories of gravity \cite{modgravWH}, electromagnetic signatures of
accretion disks around wormhole spacetimes \cite{Harko:2009xf}, etc, (we refer
the reader to \cite{Lobo:2007zb} for a review). The semi-classical procedure
followed in this work relies heavily on the formalism outlined in Ref.
\cite{Garattini}. Rather than reproduce the formalism, we shall refer the
reader to Ref. \cite{Garattini} for details, when necessary.

In this work, we explore an alternative approach to the semi-classical
approach outlined above. Note that the traditional manner is to fix a
background metric and obtain self-consistent solutions. Here, we let the
quantum fluctuations evolve, and the classical energy, which depends on the
evolved metric functions, is then evaluated. A natural ultraviolet (UV) cutoff
is obtained which forbids an interior spacetime region, and may result in a
multipy-connected spacetime. Thus, in the context of Gravity's Rainbow, this
process may be interpreted as a change in topology, and consequently results
in the presence of a Planckian wormhole.

This paper is organized in the following manner: In Section \ref{Sec1}, the
semi-classical approach is briefly outlined, and the graviton one loop
contribution to a classical energy is computed through a variational approach
with Gaussian trial wave functionals. In Section \ref{Sec2}, the
self-sustained equation is interpreted in a novel way, where the quantum
fluctuations are let to evolve and the classical energy is then computed,
consequently one arrives at solutions which may be interpreted as a change in
topology. Finally, in Section \ref{SecConc}, we conclude.

\section{The classical term and the one loop energy in Rainbow's Gravity}

\label{Sec1}

\subsection{Effective field equations in a spherically symmetric background}

In this paper, using a semi-classical approach, we explore the possibility to
directly compute a topology change and in particular the birth of a
traversable wormhole. The starting point is represented by the semiclassical
gravitational Einstein field equation given by%
\begin{equation}
G_{\mu\nu}=\kappa\left\langle T_{\mu\nu}\right\rangle ^{\mathrm{ren}},
\label{Gmunus}%
\end{equation}
where $\left\langle T_{\mu\nu}\right\rangle ^{\mathrm{ren}}$ is the
renormalized expectation value of the stress-energy tensor operator of the
quantized field, $G_{\mu\nu}$ is the Einstein tensor and $\kappa=8\pi G$.

The semi-classical procedure followed in this work relies heavily on the
formalism outlined in Ref. \cite{Garattini}, where the graviton one loop
contribution to a classical energy in a traversable wormhole background was
computed, through a variational approach with Gaussian trial wave functionals
\cite{Garattini,Garattini2}. (Note that our approach is very close to the
gravitational \textit{geon} considered by Anderson and Brill \cite{geons4b},
where the relevant difference lies in the averaging procedure). More
specifically, the metric may be separated into a background component,
$\bar{g}_{\mu\nu}$ and a perturbation $h_{\mu\nu}$, i.e., $g_{\mu\nu}=\bar
{g}_{\mu\nu}+h_{\mu\nu}$. The Einstein tensor may also be separated into a
part describing the curvature due to the background geometry and that due to
the perturbation, i.e., $G_{\mu\nu}(g_{\alpha\beta})=G_{\mu\nu}(\bar
{g}_{\alpha\beta})+\Delta G_{\mu\nu}(\bar{g}_{\alpha\beta},h_{\alpha\beta})$,
where $\Delta G_{\mu\nu}(\bar{g}_{\alpha\beta},h_{\alpha\beta})$ may be
considered a perturbation series in terms of $h_{\mu\nu}$. If the matter field
source is absent, one may define an effective stress-energy tensor for the
fluctuations as%
\begin{equation}
\left\langle T_{\mu\nu}\right\rangle ^{\mathrm{ren}}=-\frac{1}{\kappa
}\left\langle \Delta G_{\mu\nu}\left(  \bar{g}_{\alpha\beta},h_{\alpha\beta
}\right)  \right\rangle ^{\mathrm{ren}}\,.
\end{equation}

From this point of view, the equation governing quantum fluctuations behaves
as a backreaction equation. If we fix our attention to the energy component of
the Einstein field equations, we need to introduce a time-like unit vector
$u^{\mu}$ such that $u^{\mu}u_{\mu}=-1$. Then the semi-classical Einstein
equations $\left(  \ref{Gmunus}\right)  $ projected on the constant time
hypersurface $\Sigma$ are given by%
\begin{equation}
\frac{1}{2\kappa}\int_{\Sigma}d^{3}x\sqrt{^{3}\bar{g}}G_{\mu\nu}\left(
\bar{g}_{\alpha\beta}\right)  u^{\mu}u^{\nu}=-\int_{\Sigma}d^{3}%
x\mathcal{H}^{\left(  0\right)  }=-\frac{1}{2\kappa}\int_{\Sigma}d^{3}%
x\sqrt{^{3}\bar{g}}\left\langle \Delta G_{\mu\nu}\left(  \bar{g}_{\alpha\beta
},h_{\alpha\beta}\right)  u^{\mu}u^{\nu}\right\rangle ^{\mathrm{ren}},
\label{inteq}%
\end{equation}
where we have integrated the projected Einstein field equations over $\Sigma$
and where%
\begin{equation}
\mathcal{H}^{\left(  0\right)  }=2\kappa G_{ijkl}\;\pi^{ij}\pi^{kl}%
-\frac{\sqrt{\bar{g}}}{2\kappa}R\,, \label{hdens}%
\end{equation}
is the background field super-hamiltonian, $G_{ijkl}$ is the DeWitt super
metric \cite{DeWitt}, and $R$ is the curvature scalar.

In a series of papers \cite{GaMa,GaLo,GAC,MagSmo}, a distortion of the
gravitational metric known as Gravity's Rainbow was introduced as a tool to
keep the UV divergences under control. Briefly, the situation is the
following: one introduces two arbitrary functions $g_{1}\left(  E/E_{P}%
\right)  $ and $g_{2}\left(  E/E_{P}\right)  $, denoted as \textit{Rainbow's
functions}, with the only assumption that%
\begin{equation}
\lim_{E/E_{P}\rightarrow0}g_{1}\left(  E/E_{P}\right)  =1\qquad\text{and}%
\qquad\lim_{E/E_{P}\rightarrow0}g_{2}\left(  E/E_{P}\right)  =1.
\end{equation}
On a general spherical symmetric metric such functions come into play in the
following manner%
\begin{equation}
ds^{2}=-N^{2}\left(  r\right)  \frac{dt^{2}}{g_{1}^{2}\left(  E/E_{P}\right)
}+\frac{dr^{2}}{\left[  1-\frac{b\left(  r\right)  }{r}\right]  g_{2}%
^{2}\left(  E/E_{P}\right)  }+\frac{r^{2}}{g_{2}^{2}\left(  E/E_{P}\right)
}\left(  d\theta^{2}+\sin^{2}\theta d\phi^{2}\right)  \,, \label{dS}%
\end{equation}
where $N(r)$ is the lapse function and $b(r)$ is denoted the shape function.

It is clear that the classical energy on the l.h.s. of Eq. $\left(
\ref{inteq}\right)  $ is modified into the following expression \cite{GaLo}%
\begin{equation}
H_{\Sigma}^{(0)}=\int_{\Sigma}\,d^{3}x\,\mathcal{H}^{(0)}=-\frac{1}{16\pi
G}\int_{\Sigma}\,d^{3}x\,\sqrt{g}\,R=-\frac{1}{2G}\int_{r_{0}}^{\infty}%
\,\frac{dr\,r^{2}}{\sqrt{1-b(r)/r}}\,\frac{b^{\prime}(r)}{r^{2}g_{2}\left(
E\right)  }\,.
\end{equation}
For simplicity, we consider $N(r)=1$ throughout this work. Note that to be a
wormhole solution the following conditions need to be satisfied at the throat
$b(r_{0})=r_{0}$ and $b^{\prime}(r_{0})<1$; the latter condition is a
consequence of the flaring-out condition of the throat, i.e., $(b-b^{\prime
}r)/b^{2} >0$ \cite{Morris}; asymptotic flatness imposes $b(r)/r\rightarrow0$
as $r\rightarrow+\infty$.

\subsection{The one loop energy in Gravity's Rainbow}

\label{p2}

Note that the r.h.s. of Eq. $\left(  \ref{inteq}\right)  $ is represented by
the fluctuations of the Einstein tensor, which in this context, are the
fluctuations of the hamiltonian which are evaluated through a variational
approach with Gaussian trial wave functionals. The divergences are treated
with the help of the Rainbow's functions avoiding therefore the use of a
regularization and renormalization procedure. We find that the total
regularized one loop energy is given by%
\begin{equation}
E^{TT}=-\frac{1}{2}\sum_{\tau}\frac{g_{1}\left(  E\right)  }{g_{2}^{2}\left(
E\right)  }\left[  \sqrt{E_{1}^{2}\left(  \tau\right)  }+\sqrt{E_{2}%
^{2}\left(  \tau\right)  }\right]  \,, \label{ETT}%
\end{equation}
where $E_{i}\left(  \tau\right)  $ are the eigenvalues of%
\begin{equation}
\left(  \tilde{\bigtriangleup}_{L\!}^{m}\!{}\tilde{h}^{\bot}\right)  _{ij}%
\!{}=\frac{E^{2}}{g_{2}^{2}\left(  E\right)  }\tilde{h}_{ij}^{\bot}\,,
\label{EE}%
\end{equation}
with the condition that $E_{i}^{2}\left(  \tau\right)  >0$, ${h}^{\bot}$ is
the traceless-transverse component of the perturbation, and $\hat
{\bigtriangleup}_{L\!}^{m}$is defined by%
\begin{equation}
\left(  \hat{\bigtriangleup}_{L\!}^{m}\!{}h^{\bot}\right)  _{ij}=\left(
\bigtriangleup_{L\!}\!{}h^{\bot}\right)  _{ij}-4R{}_{i}^{k}\!{}h_{kj}^{\bot
}+\text{ }^{3}R{}\!{}h_{ij}^{\bot}\,. \label{M Lichn}%
\end{equation}
The operator $\bigtriangleup_{L}$ is the Lichnerowicz operator which is given
by%
\begin{equation}
\left(  \bigtriangleup_{L}h\right)  _{ij}=\bigtriangleup h_{ij}-2R_{ikjl}%
h^{kl}+R_{ik}h_{j}^{k}+R_{jk}h_{i}^{k}, \label{DeltaL}%
\end{equation}
with $\bigtriangleup=-\nabla^{a}\nabla_{a}$. We refer the reader to Ref.
\cite{Garattini} for further details.


With the help of the Regge--Wheeler representation \cite{Regge Wheeler}, the
eigenvalue equation $\left(  \ref{EE}\right)  $ can be reduced to%
\begin{equation}
\left[  -\frac{d^{2}}{dx^{2}}+\frac{l\left(  l+1\right)  }{r^{2}}+m_{i}%
^{2}\left(  r\right)  \right]  f_{i}\left(  x\right)  =\frac{E_{i,l}^{2}%
}{g_{2}^{2}\left(  E\right)  }f_{i}\left(  x\right)  \,, \label{p34}%
\end{equation}
with $i=1,2$. In Eq. $\left(  \ref{p34}\right)  $ we have used reduced fields
of the form $f_{i}\left(  x\right)  =F_{i}\left(  x\right)  /r$ and defined
two $r$-dependent effective masses $m_{1}^{2}\left(  r\right)  $ and
$m_{2}^{2}\left(  r\right)  $%
\begin{equation}
\left\{
\begin{array}
[c]{c}%
m_{1}^{2}\left(  r\right)  =\frac{6}{r^{2}}\left[  1-\frac{b(r)}{r}\right]
-\frac{3b^{\prime}(r)}{2r^{2}}+\frac{3b(r)}{2r^{3}}\,\\
\\
m_{2}^{2}\left(  r\right)  =\frac{6}{r^{2}}\left[  1-\frac{b(r)}{r}\right]
-\frac{b^{\prime}(r)}{2r^{2}}-\frac{3b(r)}{2r^{3}}%
\end{array}
\right.  \quad\left(  r\equiv r\left(  x\right)  \right)  , \label{masses}%
\end{equation}
where we have implicitly defined the variable $x$ with the help of the
following relationship $dx=dr/\sqrt{1-b(r)/r}$.

In order to use the WKB approximation, from Eq. $\left(  \ref{p34}\right)  $
we can extract two $r-$dependent radial wave numbers%
\begin{equation}
k_{i}^{2}\left(  r,l,\omega_{i,nl}\right)  =\frac{E_{i,nl}^{2}}{g_{2}%
^{2}\left(  E\right)  }-\frac{l\left(  l+1\right)  }{r^{2}}-m_{i}^{2}\left(
r\right)  \quad i=1,2\,. \label{kTT}%
\end{equation}
It is useful to use the WKB method implemented by `t Hooft in the brick wall
problem \cite{tHooft}, by counting the number of modes with frequency less
than $\omega_{i}$, $i=1,2$. This is given approximately by%
\begin{equation}
\tilde{g}\left(  E_{i}\right)  =\int_{0}^{l_{\max}}\nu_{i}\left(
l,E_{i}\right)  \left(  2l+1\right)  dl, \label{p41}%
\end{equation}
where $\nu_{i}\left(  l,E_{i}\right)  $, $i=1,2$ is the number of nodes in the
mode with $\left(  l,E_{i}\right)  $, such that $\left(  r\equiv r\left(
x\right)  \right)  $%
\begin{equation}
\nu_{i}\left(  l,E_{i}\right)  =\frac{1}{\pi}\int_{-\infty}^{+\infty}%
dx\sqrt{k_{i}^{2}\left(  r,l,E_{i}\right)  }. \label{p42}%
\end{equation}
The integration with respect to $x$ and $l$ is taken over those values which
satisfy $k_{i}^{2}\left(  r,l,E_{i}\right)  \geq0,$ $i=1,2$. With the help of
Eqs. $\left(  \ref{p41}\right)  $ and $\left(  \ref{p42}\right)  $, the
self-sustained equation becomes%
\begin{equation}
H_{\Sigma}^{(0)}=-\frac{1}{\pi}\sum_{i=1}^{2}\int_{0}^{+\infty}E_{i}%
\frac{g_{1}\left(  E\right)  }{g_{2}^{2}\left(  E\right)  }\frac{d\tilde
{g}\left(  E_{i}\right)  }{dE_{i}}dE_{i}. \label{tot1loop}%
\end{equation}
The explicit evaluation of the density of states yields%
\begin{align}
\frac{d\tilde{g}(E_{i})}{dE_{i}}  &  =\int\frac{\partial\nu(l{,}E_{i}%
)}{\partial E_{i}}(2l+1)dl\nonumber\\
&  =\frac{4}{3\pi}\int_{-\infty}^{+\infty}dxr^{2}\frac{d}{dE_{i}}\left(
\frac{E_{i}^{2}}{g_{2}^{2}\left(  E\right)  }-m_{i}^{2}\left(  r\right)
\right)  ^{\frac{3}{2}}. \label{states}%
\end{align}

Substituting Eq. $\left(  \ref{states}\right)  $ into Eq. $\left(
\ref{tot1loop}\right)  $ and taking into account the energy density, we obtain%
\begin{equation}
\frac{1}{2G}\,\frac{b^{\prime}(r)}{r^{2}g_{2}\left(  E\right)  }=\frac{2}%
{3\pi^{2}}\left(  I_{1}+I_{2}\right)  , \label{LoverG}%
\end{equation}
where the integrals $I_{1}$ and $I_{2}$ are respectively given by%
\begin{equation}
I_{1}=\int_{E^{\ast}}^{\infty}E\frac{g_{1}\left(  E\right)  }{g_{2}^{2}\left(
E\right)  }\frac{d}{dE}\left[  \frac{E^{2}}{g_{2}^{2}\left(  E\right)  }%
-m_{1}^{2}\left(  r\right)  \right]  ^{\frac{3}{2}}dE=3\int_{E^{\ast}}%
^{\infty}E^{2}\frac{g_{1}\left(  E\right)  }{g_{2}^{3}\left(  E\right)  }%
\sqrt{\frac{E^{2}}{g_{2}^{2}\left(  E\right)  }-m_{1}^{2}\left(  r\right)
}\frac{d}{dE}\left(  \frac{E}{g_{2}\left(  E\right)  }\right)  dE\,,
\label{I1}%
\end{equation}
and%
\begin{equation}
I_{2}=\int_{E^{\ast}}^{\infty}E\frac{g_{1}\left(  E\right)  }{g_{2}^{2}\left(
E\right)  }\frac{d}{dE}\left[  \frac{E^{2}}{g_{2}^{2}\left(  E\right)  }%
-m_{2}^{2}\left(  r\right)  \right]  ^{\frac{3}{2}}dE=3\int_{E^{\ast}}%
^{\infty}E^{2}\frac{g_{1}\left(  E\right)  }{g_{2}^{3}\left(  E\right)  }%
\sqrt{\frac{E^{2}}{g_{2}^{2}\left(  E\right)  }-m_{2}^{2}\left(  r\right)
}\frac{d}{dE}\left(  \frac{E}{g_{2}\left(  E\right)  }\right)  dE\,,
\label{I2}%
\end{equation}
where $E^{\ast}$ is the value which annihilates the argument of the root. In
$I_{1}$ and $I_{2}$ we have assumed that the effective mass does not depend on
the energy $E$. The purpose of this paper is to show that the self-sustained
equation $\left(  \ref{LoverG}\right)  $ is also a source of a topology change.

\section{Topology change}

\label{Sec2}

Now, Eq. $\left(  \ref{LoverG}\right)  $ can be read off in a twofold way. The
traditional manner is to fix the same background on both sides and verify the
existence of a consistent solution on the remaining parameters, e.g., the
wormhole throat, establishing the existence of a self sustained traversable
wormhole \cite{Garattini,Garattini:2007ff}. However, one may also adopt an
alternative approach, where one fixes the background on the r.h.s. of Eq.
$\left(  \ref{LoverG}\right)  $ and consequently let the quantum fluctuations
evolve, and then one verifies what kind of solutions we can extract from the
l.h.s. in a recursive way. To fix ideas, Eq. $\left(  \ref{LoverG}\right)  $
should be read off in the following manner%
\begin{equation}
\frac{1}{2G}\,\frac{\left(  b^{\prime}(r)\right)  ^{\left(  n\right)  }}%
{r^{2}g_{2}\left(  E\right)  }=\frac{2}{3\pi^{2}}\left[  I_{1}\left(
b^{\left(  n-1\right)  }(r)\right)  +I_{2}\left(  b^{\left(  n-1\right)
}(r)\right)  \right]  , \label{LoG}%
\end{equation}
where $n$ is the order of the approximation. In this way, if we discover that
the l.h.s. has solutions which topologically differ from the fixed background
of the r.h.s., we can conclude that a topology change has been induced from
quantum fluctuations of the graviton for any spherically symmetric background
on the r.h.s of Eq. $\left(  \ref{LoverG}\right)  $. Of course, it is not a
trivial task to realize multiple topology changes, even if Eq. $\left(
\ref{LoG}\right)  $ is interpreted in this way.

The simplest way to see if Eq. $\left(  \ref{LoG}\right)  $ allows a topology
change is to fix the Minkowski background on the r.h.s. This means that
$b(r)=0$ $\forall r$ and for $n=1$, so that the r.h.s. of Eq. $\left(
\ref{LoG}\right)  $ reduces to%
\begin{equation}
\frac{1}{2G}\,\frac{b^{\prime}(r)}{r^{2}g_{2}\left(  E\right)  }=\frac{4}%
{\pi^{2}}\int_{E^{\ast}}^{\infty}E^{2}\frac{g_{1}\left(  E\right)  }{g_{2}%
^{3}\left(  E\right)  }\sqrt{\frac{E^{2}}{g_{2}^{2}\left(  E\right)  }%
-\frac{6}{r^{2}}}\frac{d}{dE}\left(  \frac{E}{g_{2}\left(  E\right)  }\right)
dE. \label{LoG1}%
\end{equation}
Then all one has to do is determine the output of the l.h.s. of Eq. $\left(
\ref{LoG1}\right)  $, namely the classical energy in Gravity's Rainbow.
Nevertheless the result is strongly dependent on the choices that we impose on
the Rainbow's functions. We will discuss two specific examples which, of
course, do not exhaust all the possible choices. Nevertheless, these specific
cases show that in principle if we adopt the alternative approach outlined
above, one arrives at a solution that can be interpreted as a change in
topology.\textbf{ }Note that Eq.$\left(  \ref{LoG}\right)  $ is strictly
related to the \textquotedblleft Ricci flow\textquotedblright\ which is a good
tool to detect and compute a topology change\cite{Dzhunushaliev,MHTW}. A Ricci
flow is defined as follows%
\begin{equation}
\frac{\partial g_{\mu\nu}\left(  x^{\rho},\lambda\right)  }{\partial\lambda
}=-2R_{\mu\nu}\left(  x^{\rho},\lambda\right)  , \label{Rf}%
\end{equation}
where $R_{\mu\nu}$ is the Ricci curvature; $\lambda$ is an evolution parameter
which has the dimension of (length)$^{2}$ and $x^{\rho}$ are the local
coordinates on a manifold $M$. The indices $\mu,\nu,\rho$ run from $1$ to
$n=\dim M$. Nothing forbids to consider only the spatial part of the metric
appearing in Eq. $\left(  \ref{Rf}\right)  $ assuming the following form%
\begin{equation}
dl^{2}=\frac{dr^{2}}{1-\frac{b\left(  r,\lambda\right)  }{r}}+r^{2}\left(
d\theta^{2}+\sin^{2}\theta d\phi^{2}\right)  \,. \label{dl}%
\end{equation}
In this way all the topology change information is encoded in the shape
function $b\left(  r,\lambda\right)  $. In order to make a connection between
the Ricci flow and Eq.$\left(  \ref{LoG}\right)  $, it is convenient the
following setting%
\begin{equation}
\lambda\rightarrow E/E_{P}\qquad R_{\mu\nu}\left(  x^{\rho},\lambda\right)
\rightarrow CGR_{\mu\nu}\left(  x^{\rho},E/E_{P}\right)  ,
\end{equation}
where $C$ is a dimensionless constant and $G$ is the Newton's constant. Thus,
by contracting Eq.$\left(  \ref{Rf}\right)  $ with $g^{ij}\left(
x^{k},E/E_{P}\right)  $, one gets%
\begin{equation}
E_{P}\frac{g^{ij}\left(  x^{k},E/E_{P}\right)  \partial g_{ij}\left(
x^{k},E/E_{P}\right)  }{\partial E}=-2R\left(  x^{k},E/E_{P}\right)
=-4\frac{b^{\prime}(r,E/E_{P})}{r^{2}}.
\end{equation}
Multiplying both sides of the previous equation by $1/\left(  8G^{2}%
g_{2}\left(  E/E_{P}\right)  \right)  $, we find%
\begin{equation}
\frac{E_{P}}{8G^{2}g_{2}\left(  E/E_{P}\right)  }\frac{g^{ij}\left(
x^{k},E/E_{P}\right)  \partial g_{ij}\left(  x^{k},E/E_{P}\right)  }{\partial
E}=-\frac{Cb^{\prime}(r,E/E_{P})}{2Gg_{2}\left(  E/E_{P}\right)  r^{2}}.
\label{Rf1}%
\end{equation}
Comparing the r.h.s. of Eq.$\left(  \ref{Rf1}\right)  $ with the l.h.s. of
Eq.$\left(  \ref{LoG}\right)  $ we can see that the one loop contribution of
the graviton in the r.h.s. of Eq.$\left(  \ref{LoG}\right)  $ can be connected
to the contracted Ricci flow. However a detailed analysis on the connection
between the Ricci flow and the iterative procedure exposed in Eq.$\left(
\ref{LoG}\right)  $ is beyond the scope of this paper. To see if a topology
change appears, in the next sections we will discuss specific examples.

\subsection{Specific example I: $g_{1}\left(  E/E_{P}\right)  =\exp
(-\alpha\frac{E^{2}}{E_{P}^{2}}),\qquad g_{2}\left(  E/E_{P}\right)  =1$}

Following Ref. \cite{GaMa}, we consider the following choice for the Rainbow's
functions
\begin{equation}
g_{1}\left(  E/E_{P}\right)  =\exp(-\alpha\frac{E^{2}}{E_{P}^{2}}),\qquad
g_{2}\left(  E/E_{P}\right)  =1;\qquad\alpha>0\in\mathbb{R}. \label{g1g2}%
\end{equation}
then the integrals $I_{1}$ and $I_{2}$ take the following form%
\begin{align}
I_{1}=I_{2}  &  =3\int_{E^{\ast}}^{\infty}\exp(-\alpha\frac{E^{2}}{E_{P}^{2}%
})E^{2}\sqrt{E^{2}-\frac{6}{r^{2}}}dE\nonumber\\
&  =\frac{3}{2}E_{P}^{4}\int_{\sqrt{6}/r}^{\infty}\exp(-\alpha x)\sqrt{x}%
\sqrt{x-\frac{6}{\left(  rE_{P}\right)  ^{2}}}dx\nonumber\\
&  =\frac{3E_{P}^{4}}{2\sqrt{\pi}}\left(  \frac{6}{\alpha\left(
rE_{P}\right)  ^{2}}\right)  \Gamma\left(  \frac{3}{2}\right)  \exp\left(
-\frac{3\alpha}{\left(  rE_{P}\right)  ^{2}}\right)  K_{1}\left(
\frac{3\alpha}{\left(  rE_{P}\right)  ^{2}}\right)  ,
\end{align}
where we have used the following change of variables $E=\sqrt{x}$ and the
following relationship
\begin{equation}
\int_{u}^{\infty}\left(  x-u\right)  ^{\mu-1}x^{\mu-1}\exp\left(  -\beta
x\right)  dx=\frac{1}{\sqrt{\pi}}\left(  \frac{u}{\beta}\right)  ^{\mu
-1/2}\Gamma\left(  \mu\right)  \exp\left(  -\frac{\beta u}{2}\right)
K_{\mu-1/2}\left(  \frac{\beta u}{2}\right)  \qquad%
\begin{array}
[c]{c}%
\mathrm{Re}\,\mu>0\\
\mathrm{Re}\,\beta u>0
\end{array}
.
\end{equation}
Equation $(\ref{LoG1})$ can be rearranged in the following way%
\begin{equation}
\frac{1}{2G}\,\frac{b^{\prime}(r)}{r^{2}}=\frac{E_{P}^{4}}{\pi^{2}}\left[
\frac{6}{\alpha\left(  rE_{P}\right)  ^{2}}\exp\left(  -\frac{3\alpha}{\left(
rE_{P}\right)  ^{2}}\right)  K_{1}\left(  \frac{3\alpha}{\left(
rE_{P}\right)  ^{2}}\right)  \right]  , \label{bp}%
\end{equation}
where $K_{1}\left(  x\right)  $ is a modified Bessel function of order 1. Note
that it is extremely difficult to extract any useful information from this
relationship, so that in the following we consider two regimes, namely the
cis-planckian regime, where $rE_{P}\gg1$ and the trans-planckian r\'{e}gime,
where $rE_{P}\ll1$.

In the cis-planckian regime, expanding the right hand side of Eq. $\left(
\ref{bp}\right)  $, we find that the leading term is given by
\begin{equation}
\frac{1}{G}\,\frac{b^{\prime}(r)}{r^{2}}\simeq\frac{E_{P}^{4}}{\pi^{2}}\left[
\frac{4}{\alpha^{2}}-\frac{12}{\alpha\left(  rE_{P}\right)  ^{2}}+O\left(
\left(  rE_{P}\right)  ^{-4}\right)  \right]  \,
\end{equation}
which can be rearranged to give%
\begin{equation}
b^{\prime}(r)=\frac{1}{\pi^{2}}\left[  \frac{4r^{2}}{\alpha^{2}G}-\frac
{12}{\alpha}\right]  \,,
\end{equation}
where we have used the definition $G=E_{P}^{-2}=l_{P}^{2}$. Restricting our
attention to the dominant term only, we find that
\begin{equation}
b(r)=r_{t}+\frac{4E_{P}^{2}}{3\pi^{2}\alpha^{2}}\left(  r^{3}-r_{t}%
^{3}\right)  -\frac{12}{\alpha}\left(  r-r_{t}\right)  \,,
\end{equation}
which does not represent an asymptotically flat wormhole geometry, as the
condition $b(r)/r\rightarrow0$ when $r\rightarrow+\infty$, is not satisfied.
On the other hand, in the trans-planckian regime, i.e., $rE_{P}\ll1$, we
obtain the following approximation
\begin{equation}
\,\frac{b^{\prime}(r)}{r^{2}}\simeq\frac{E_{P}^{2}}{2\sqrt{\alpha^{3}\pi^{3}}%
}\left[  \exp\left(  -\alpha\frac{6}{\left(  rE_{P}\right)  ^{2}}\right)
\frac{\sqrt{6}}{rE_{P}}+O\left(  rE_{P}\right)  \right]  \,.
\end{equation}
Note that in this regime, the asymptotic expansion is dominated by the
Gaussian exponential so that the quantum correction vanishes. Thus, the only
solution is $b^{\prime}(r)=0$ and consequently we have a constant shape
function, namely, $b(r)=r_{t}$. In the next two examples, we will consider
different forms of the Rainbow's functions, in order to verify if a topology
change may occur.

\subsection{Specific example II: $g_{1}(E/E_{P})=g_{2}(E/E_{P})=g(E/E_{P})$}

Reintroducing the Planck scale $E_{P}$ explicitly, we consider the specific
choice%
\begin{equation}
g_{1}(E/E_{P})=g_{2}(E/E_{P})=g(E/E_{P})\,,
\end{equation}
then the integrals $I_{1}$ and $I_{2}$ take the following form%
\begin{align}
I_{1}=I_{2}  &  =3\int_{E^{\ast}}^{\infty}\left(  \frac{E}{g_{2}\left(
E/E_{P}\right)  }\right)  ^{2}\sqrt{\left(  \frac{E}{g_{2}\left(
E/E_{P}\right)  }\right)  ^{2}-\frac{6}{r^{2}}}\frac{d}{dE}\left(  \frac
{E}{g_{2}\left(  E/E_{P}\right)  }\right)  dE\nonumber\\
&  =3E_{P}^{4}\int_{\sqrt{6}/r}^{x_{\infty}}x^{2}\sqrt{x^{2}-\frac{6}{r^{2}}%
}dx\,,
\end{align}
and consequently Eq. $\left(  \ref{LoG1}\right)  $ simplifies to%
\begin{equation}
\,\frac{b(r)}{g\left(  E/E_{P}\right)  }=\frac{8GE_{P}^{4}}{\pi^{2}}%
\int\left[  \int_{\sqrt{6}/r^{\prime}}^{x_{\infty}}x^{2}\sqrt{x^{2}-\frac
{6}{r^{^{\prime}2}}}dx\right]  r^{\prime2}dr^{\prime}+C, \label{top1}%
\end{equation}
where $C$ is an arbitrary constant fixed by boundary conditions and where we
have defined the following parameters%
\begin{equation}
x=\frac{E/E_{P}}{g\left(  E/E_{P}\right)  }\qquad\mathrm{and}\qquad x_{\infty
}=\lim_{E/E_{P}\rightarrow\infty}\frac{E/E_{P}}{g\left(  E/E_{P}\right)  }\,.
\end{equation}
The integration inside the square brackets, in Eq. $\left(  \ref{top1}\right)
$, is straightforward to calculate and one finally arrives at $\left(
G=E_{P}^{-2}\right)  $%
\begin{align}
\frac{b(r)}{g\left(  E/E_{P}\right)  }  &  =\frac{2E_{P}^{2}}{\pi^{2}}%
\int\Bigg[x_{\infty}\sqrt{\left(  x_{\infty}^{2}-\frac{6}{\left(
E_{P}r^{\prime}\right)  ^{2}}\right)  ^{3}}+\frac{6x_{\infty}}{\left(
E_{P}r^{\prime}\right)  ^{2}}\sqrt{x_{\infty}^{2}-\frac{6}{\left(
E_{P}r^{\prime}\right)  ^{2}}}\nonumber\\
&  -\frac{18}{\left(  E_{P}r^{\prime}\right)  ^{4}}\ln\left(  x_{\infty}%
+\sqrt{x_{\infty}^{2}-\frac{6}{\left(  E_{P}r^{\prime}\right)  ^{2}}}\right)
+\frac{18}{\left(  E_{P}r^{\prime}\right)  ^{4}}\ln\left(  \sqrt{\frac
{6}{\left(  E_{P}r^{\prime}\right)  ^{2}}}\right)  \Bigg]r^{\prime2}%
dr^{\prime}+C. \label{b(r)}%
\end{align}

Now, we have to fix the form of $g(E/E_{P})$. As a working hypothesis,
consider%
\begin{equation}
g(E/E_{P})=1+\left(  E/E_{P}\right)  ^{\alpha}%
\end{equation}
with $\alpha>0$, where the factor $x_{\infty}$ is given by%
\begin{equation}
x_{\infty}=\lim_{E/E_{P}\rightarrow\infty}\frac{E/E_{P}}{g\left(
E/E_{P}\right)  }=\left\{
\begin{array}
[c]{cc}%
+\infty & \alpha<1\\
1 & \alpha=1\\
0 & \alpha>1
\end{array}
\right.  . \label{lim}%
\end{equation}
Due to the divergence present in $\alpha<1$, and as $\alpha>1$ leads to an
imaginary result, the only possible choice is given by $\alpha=1$. Note that
the case under consideration has been extensively studied in inflationary
cosmology \cite{alpha}. By using the explicit form of $g(E/E_{P})$, Eq.
$\left(  \ref{b(r)}\right)  $ reduces to%
\begin{equation}
\frac{b(r)}{1+E/E_{P}}=\frac{2E_{P}^{2}}{\pi^{2}}\int h\left(  r^{\prime
}\right)  r^{\prime2}dr^{\prime}\,+C, \label{b(r)split}%
\end{equation}
where%
\begin{equation}
\,\,h\left(  r^{\prime}\right)  =\frac{1}{\left(  E_{P}r^{\prime}\right)
^{3}}\left\{  \left[  \left(  E_{P}r^{\prime}\right)  ^{2}-6\right]
^{\frac{3}{2}}+6\left[  \left(  E_{P}r^{\prime}\right)  ^{2}-6\right]
^{\frac{1}{2}}-\frac{18}{E_{P}r^{\prime}}\ln\left[  \frac{E_{P}r^{\prime}%
}{\sqrt{6}}+\left(  \frac{\left(  E_{P}r^{\prime}\right)  ^{2}}{6}-1\right)
^{\frac{1}{2}}\right]  \right\}  \,. \label{h(r)}%
\end{equation}
It is immediate to observe that there is a natural UV cutoff which forbids the
use of an interior region $[0,\sqrt{6}/E_{P}]$. Indeed, below the point
$r=\sqrt{6}/E_{P}$, the integrand becomes imaginary. The integration in the
interval $\left[  0,\sqrt{6}/E_{P}\right]  $ can be thought as a
trans-planckian contribution to the shape function which is also suppressed by
a factor $E/E_{P}$. On the other hand in the cis-planckian region,
$g(E/E_{P})\simeq1$ and%
\begin{equation}
b(r)=\frac{2E_{P}^{2}}{\pi^{2}}\int h\left(  r^{\prime}\right)  r^{\prime
2}dr^{\prime}\,+C.
\end{equation}

Thus, the natural UV cutoff which forbids the use of an interior region, may
be interpreted as a topology change. Now we have to verify if this change has
produced a wormhole solution. The throat condition is satisfied by definition
if one imposes that $b(r_{0})=r_{0}=$ $\sqrt{6}/E_{P}=C$. When $r^{\prime}%
\gg\sqrt{6}/E_{P}$, we can write%
\begin{equation}
h\left(  r^{\prime}\right)  \simeq1-\frac{3}{\left(  E_{P}r^{\prime}\right)
^{2}}-\frac{9}{2\left(  E_{P}r^{\prime}\right)  ^{4}}. \label{h(r)as}%
\end{equation}
and one finds%
\begin{align}
b(r)  &  =\frac{2E_{P}^{2}}{\pi^{2}}\int_{r_{0}}^{r}\left(  1-\frac{3}{\left(
E_{P}r^{\prime}\right)  ^{2}}-\frac{9}{2\left(  E_{P}r^{\prime}\right)  ^{4}%
}\right)  r^{\prime2}dr^{\prime}\,+r_{0}\nonumber\\
&  =\frac{2E_{P}^{2}}{\pi^{2}}\left[  \frac{r^{3}-r_{0}^{3}}{3}-\frac
{3}{{E_{{P}}^{2}}}\left(  {r-}r_{0}\right)  +\frac{9}{2{E_{{P}}^{4}}}\left(
\frac{1}{r}-\frac{1}{r_{0}}\right)  \right]  +r_{0}.
\end{align}

To be a wormhole solution, the flaring-out condition at the throat, i.e.,
$b^{\prime}(r_{0})<1$, needs to be obeyed. The radial derivative of the shape
function is given by $b^{\prime}(r)=\frac{12}{\pi^{2}}\left[  \left(  \frac
{r}{r_{0}} \right)  ^{2} -\frac{1}{2} -\frac{1}{8}\left(  \frac{r_{0}}%
{r}\right)  ^{2} \right]  $, which reduces to $b^{\prime}(r_{0})=\frac{9}%
{2\pi^{2}} $ at the throat, so that the flaring-out condition is satisfied.
Note that the resulting space is not asymptotically flat, but rather
asymptotically de Sitter; therefore the condition $b(r)/r\rightarrow0$ when
$r\rightarrow\infty$ is not satisfied. However, in principle, one may match
this interior solution to an exterior vacuum much in the spirit of Refs.
\cite{match}.

\subsection{Specific example III: $g_{2}\left(  E/E_{P}\right)  =1+E/E_{P}$
and $g_{1}\left(  E/E_{P}\right)  =g\left(  E/E_{P}\right)  \left(
1+E/E_{P}\right)  ^{6}$}

A second interesting example can be proposed starting from Minkowski space
with the choice%
\begin{equation}
g_{2}\left(  E/E_{P}\right)  =1+E/E_{P}\,,
\end{equation}
which is necessary in order to avoid divergent or imaginary values of the
integral. Thus, we obtain $I_{1}=I_{2}=I$, and the integral $I$ is given by%
\begin{align}
I  &  =E_{P}^{2}\int_{E^{\ast}}^{\infty}E^{2}\frac{g_{1}\left(  E/E_{P}%
\right)  }{\left(  1+E/E_{P}\right)  ^{3}}\sqrt{\left(  \frac{E/E_{P}%
}{1+E/E_{P}}\right)  ^{2}-\frac{6}{\left(  E_{P}r\right)  ^{2}}}\;\frac{d}%
{dE}\left(  \frac{E/E_{P}}{1+E/E_{P}}\right)  dE\nonumber\\
&  =E_{P}^{4}\int_{\frac{E_{P}\sqrt{6}}{rE_{P}-\sqrt{6}}}^{\infty}g\left(
\frac{E}{E_{P}}\right)  \left(  1+\frac{E}{E_{P}}\right)  \left(  \frac
{E}{E_{P}}\right)  ^{2}\sqrt{\left(  \frac{E/E_{P}}{1+E/E_{P}}\right)
^{2}-\frac{6}{\left(  E_{P}r\right)  ^{2}}}\;d\left(  \frac{E}{E_{P}}\right)
,
\end{align}
where we have set%
\begin{equation}
g_{1}\left(  \frac{E}{E_{P}}\right)  =g\left(  \frac{E}{E_{P}}\right)  \left(
1+\frac{E}{E_{P}}\right)  ^{6}\,,
\end{equation}
and inserted explicitly the value of $E^{\ast}$. The term $rE_{P}$ becomes
relevant when $r\sim E_{P}^{-1}$ and integrating over $r$ one gets%
\begin{equation}
\frac{b(r)}{1+E/E_{P}}=\frac{2E_{P}^{2}}{\pi^{2}}\int\left[  \int_{\frac
{E_{P}\sqrt{6}}{r^{\prime}E_{P}-\sqrt{6}}}^{\infty}g\left(  \frac{E}{E_{P}%
}\right)  \left(  1+\frac{E}{E_{P}}\right)  \left(  \frac{E}{E_{P}}\right)
^{2}\sqrt{\left(  \frac{E/E_{P}}{1+E/E_{P}}\right)  ^{2}-\frac{6}{\left(
E_{P}r^{\prime}\right)  ^{2}}}d\left(  \frac{E}{E_{P}}\right)  \right]
r^{\prime2}dr^{\prime}\,+C.
\end{equation}

In this case, we also have a short distance cut off located at $r^{\prime
}=\sqrt{6}/E_{P}$, below which the energy $E$ becomes negative and the
argument under square root becomes imaginary. Thus, by repeating the same
steps we did for the example I, we find that the integration over the first
interval can be interpreted as a trans-planckian region and therefore also
suppressed by a factor $E/E_{P}$. Thus, we remain with%
\begin{equation}
b(r)=\frac{2E_{P}^{2}}{\pi^{2}}\int\left[  \int_{\frac{E_{P}\sqrt{6}%
}{r^{\prime}E_{P}-\sqrt{6}}}^{\infty}g\left(  \frac{E}{E_{P}}\right)  \left(
1+\frac{E}{E_{P}}\right)  \left(  \frac{E}{E_{P}}\right)  ^{2}\sqrt{\left(
\frac{E/E_{P}}{1+E/E_{P}}\right)  ^{2}-\frac{6}{\left(  E_{P}r^{\prime
}\right)  ^{2}}}d\left(  \frac{E}{E_{P}}\right)  \right]  r^{\prime
2}dr^{\prime}\,+C \label{b(r)g}%
\end{equation}
Assuming that the throat is located at $r_{0}=\sqrt{6}/E_{P}$, the condition
$b(r_{0})=r_{0}=C$ is automatically satisfied. It is important to observe that
the factor $6/\left(  E_{P}r^{\prime}\right)  ^{2}$ is highly suppressed in
the region $\left[  \sqrt{6}/E_{P},r\right]  $. Therefore Eq. $\left(
\ref{b(r)g}\right)  $ can be approximated to give%
\begin{equation}
b(r)=\frac{2E_{P}^{2}}{\pi^{2}}\int_{r_{0}}^{r}\left[  \int_{\frac{\sqrt{6}%
}{r^{\prime}E_{P}}}^{\infty}g\left(  \frac{E}{E_{P}}\right)  \left(  \frac
{E}{E_{P}}\right)  ^{3}d\left(  \frac{E}{E_{P}}\right)  \right]  r^{\prime
2}dr^{\prime}\,+r_{0}\,.
\end{equation}

For simplicity, fixing $g\left(  E/E_{P}\right)  =\exp\left(  -\alpha
E/E_{P}\right)  \left(  1+\beta E/E_{P}\right)  $ with $\beta\in\mathbb{R}$,
one obtains%
\begin{align}
b(r)  &  =\frac{2E_{P}^{2}}{\pi^{2}}\int_{r_{0}}^{r}\left[  -\frac{d^{3}%
}{d\alpha^{3}}\int_{\frac{\sqrt{6}}{r^{\prime}E_{P}}}^{\infty}\exp\left(
-\alpha E/E_{P}\right)  d\left(  \frac{E}{E_{P}}\right)  +\beta\frac{d^{4}%
}{d\alpha^{4}}\int_{\frac{\sqrt{6}}{r^{\prime}E_{P}}}^{\infty}\exp\left(
-\alpha E/E_{P}\right)  d\left(  \frac{E}{E_{P}}\right)  \right]  r^{\prime
2}dr^{\prime}\,+r_{0}\nonumber\\
&  =\frac{12E_{P}^{2}}{\pi^{2}\alpha}\int_{r_{0}}^{r}\exp\left(  -\frac
{\alpha\sqrt{6}}{E_{{P}}r^{\prime}}\right)  h(\alpha,\beta,r^{\prime
})dr^{\prime}+r_{0},
\end{align}
where%
\begin{equation}
h(\alpha,\beta,r^{\prime})=\frac{1}{{\alpha}^{3}}+\frac{\sqrt{6}}{E{_{{P}}%
}r^{\prime}{\alpha}^{2}}+\frac{3}{{E_{{P}}^{2}}r^{\prime2}{\alpha}}%
+\frac{\sqrt{6}}{{E_{{P}}^{3}}r^{\prime3}}+\beta\left(  \frac{6}{{E_{{P}}^{4}%
}r^{\prime4}}+\frac{4\sqrt{6}}{r^{\prime3}{\alpha E_{{P}}^{3}}}+\frac
{12}{r^{\prime2}{\alpha}^{2}{E_{{P}}^{2}}}+\frac{4\sqrt{6}}{r^{\prime}{\alpha
}^{3}{E_{{P}}}}+\frac{4}{{\alpha}^{4}}\right)  .
\end{equation}
After the integration over $r^{\prime}$, for large values of $r$, one obtains%
\begin{align}
b(r)  &  =r_{0}+\frac{2E_{P}^{2}}{\pi^{2}}\left[  \frac{2}{{\alpha}^{5}%
}\left(  {4\beta}+{\alpha}\right)  r^{3}-\frac{2\sqrt{6}}{{e}^{\alpha}{E_{{P}%
}^{3}\alpha}^{5}}\left[  \left(  {{\alpha}^{3}}\left(  3+{{e}^{\alpha}%
}\right)  +12\left(  2{+2{\alpha+\alpha}^{2}}\right)  \right)  \beta+\left(
{{\alpha}^{3}}\left(  3+{{e}^{\alpha}{{\alpha}}}\right)  {+6{{\alpha}}}\left(
1{+{\alpha}}\right)  \right)  \right]  \right. \nonumber\\
&  +\left.  {\frac{9}{{E_{{P}}^{4}}r}}+{\frac{18\sqrt{6}}{5r^{2}{E_{{P}}^{5}}%
}}\,\left(  \beta\,-\alpha\right)  \right]  +O\left(  {r}^{-3}\right)  .
\end{align}
Contrary to the previous case, when we set $\beta=-\alpha/4$, the de Sitter
behavior is eliminated. Plugging such a choice of parameters into the shape
function, one finds the following asymptotically flat solution%
\begin{equation}
b(r)=r_{0}+\frac{3\sqrt{6}}{\pi^{2}E_{P}{{e}^{\alpha}}\alpha}\left\{
1-\exp\left[  \left(  1-\frac{\sqrt{6}}{E_{{P}}r}\right)  \alpha\right]
\right\}  \,. \label{TCW}%
\end{equation}
The resulting space is asymptotically flat, i.e., $b(r)/r\rightarrow0$ when
$r\rightarrow\infty$ is satisfied. The radial derivative of the shape function
is given by $b^{\prime}(r)=-\frac{3r_{0}^{2}}{\pi e^{\alpha}r^{2}}\exp\left[
\left(  1-\frac{r_{0}}{r}\right)  \alpha\right]  $. Thus, the latter evaluated
at the throat reduces $b^{\prime}(r_{0})=-3/(\pi e^{\alpha})$, which satisfies
the flaring-out condition at the throat, i.e., $b^{\prime}(r_{0})<1$ for all
values of $\alpha$. One can observe that the topology change could have been
approached also distorting the one loop graviton by means of a Noncommutative
geometry like in Ref.\cite{RGPN,RemoTM,RemoKSM}, where the classical Liouville
measure has been modified into\cite{RGPN}%
\begin{equation}
dn_{i}=\frac{d^{3}\vec{x}d^{3}\vec{k}}{\left(  2\pi\right)  ^{3}}\exp\left(
-\frac{\theta}{4}\left(  \omega_{i,nl}^{2}-m_{i}^{2}\left(  r\right)  \right)
\right)  ,\quad i=1,2. \label{moddn}%
\end{equation}
$m_{i}^{2}\left(  r\right)  $ are the effective masses described in $\left(
\ref{masses}\right)  $ and $\theta$ is the Noncommutative parameter. What it
has been obtained is that the usage of the distorted Liouville measure
$\left(  \ref{moddn}\right)  $ produces a wormhole which is traversable in
principle but not in practice, but from the topology change point of view it
is immediate to see that a Noncommutative approach is less flexible with
respect to Gravity's Rainbow because of the $\theta$ parameter. On this ground
one could be interested to see how different distortions can produce a
topology change like in a Ho\v{r}ava-Lifshitz (HL) theory. Needless to say
that in a HL theory, the first step is to see if traversable wormholes are
allowed in the more general setting of a HL theory without detailed balanced
condition. In this framework the large number of coupling constant allows a
certain degree of flexibility which is actually under
investigation\cite{RGFSNLHL}.

\section{Summary and discussion}

\label{SecConc} \label{p4}

In this paper we have studied the problem of topology change induced by vacuum
fluctuations on a fixed background. The calculation has been realized
computing the graviton one-loop contribution in a semi-classical approach. The
method is based on a variational approach with Gaussian trial wave functionals
which is closely related to the Feynman path integral approach. From this
point of view, our result is not in contrast with the Hawking conjecture
forbidding a topology change. It is interesting to note that the usual UV
divergences are kept under control using a distortion of the usual
gravitational background metric known as Gravity's Rainbow which activates at
Planck's scale where it is supposed that the structure of spacetime begins to
become foamy. In practice, the energy density of the graviton one-loop
contribution, or equivalently the background spacetime, is let to evolve, and
consequently the classical energy is determined. More specifically, the
background metric is fixed to be Minkowskian in the equation governing the
quantum fluctuations, which behaves essentially as a backreaction equation,
and the quantum fluctuations are let to evolve. The classical energy, which
depends on the evolved metric functions, is then evaluated. Analysing this
procedure, a natural UV cutoff is obtained, which forbids the presence of an
interior spacetime region, and may result in a multipy-connected spacetime.
Note that one could interpret the UV cutoff as a failure of the WKB
approximation because the interior region becomes imaginary. However this
appear only in examples II and III, where the Rainbow's functions do not
quickly eliminate the divergent behavior as the example I shows. It is
important also to remark that the validity range of the WKB approximation is
in the high energy sector which is well satisfied by the trans-planckian
regime which we are probing. However, let us assume that one must include the
interior region producing an imaginary contribution. From the imposition that
$b\left(  r\right)  $ must be real, this imaginary contribution is
automatically cut off. Moreover, let us suppose that the Gravity's Rainbow
argument does not apply to this case, then to have finite results one could
impose a UV cutoff by hand in Eq. $\left(  \ref{LoG1}\right)  $ to have finite
results. This UV cutoff simply becomes%
\begin{equation}
r\geq\frac{\sqrt{6}}{\Lambda_{UV}},
\end{equation}
which means that one cannot impose the Minkowski space as a reference space
and therefore there is no a topology change. However Gravity's Rainbow allows
to consider $\Lambda_{UV}\rightarrow\infty$. This means that the previous
inequality simply reduces to $r\geq0$, namely Minkowski space. Thus, in the
context of Gravity's Rainbow, this process may be interpreted as a change in
topology, and consequently results in the presence of a Planckian wormhole.

Note that in principle, one can adopt other backgrounds including a positive
cosmological constant, i.e., a de Sitter spacetime, or a negative cosmological
constant, namely an Anti-de Sitter spacetime concluding that a hole can be
produced by Zero Point Energy quantum fluctuations. Finally using the Casimir
energy indicator \cite{DBGL}, one can conclude that the presence of holes in
spacetime seems to be favored leading therefore to a multiply connected
spacetime. It is also interesting to note that one could compute the
transition from Minkowski to a de Sitter spacetime or Anti-de Sitter
spacetime. It is important to remark that once the topology has been changed
nothing can be said on the classical/quantum stability of the new spacetime
because the whole calculation has been performed without a time evolution. It
would also be interesting to consider solutions with charge. Indeed in
\cite{Garattini:2008qs}, the Wheeler--DeWitt equation was considered as a
device for finding eigenvalues of a Sturm Liouville problem. In particular,
the Maxwell charge was interpreted as an eigenvalue of the Wheeler--De Witt
equation generated by the gravitational field fluctuations. More specifically,
it was shown that electric/magnetic charges could be generated by quantum
fluctuations of the pure gravitational field. It would also be interesting to
consider the presence of electric/magnetic charges in the solutions outlined
in this paper, and work along these lines is currently underway.

\section*{Acknowledgements}

FSNL acknowledges financial support of the Funda\c{c}\~{a}o para a Ci\^{e}ncia
e Tecnologia through the grants CERN/FP/123615/2011 and CERN/FP/123618/2011.

\end{document}